# Nano scale thermo-electrical detection of magnetic domain wall propagation

**Authors**

Patryk Krzysteczko[1], James Wells[2], Alexander Fernandez Scarioni[1], Zbynek Soban[3], Tomas Janda[3,4], Xiukun Hu[1], Vit Saidl[4], Richard P. Campion[6], Rhodri Mansell[7], Ji-Hyun Lee[7], Russell P. Cowburn[7], Petr Nemec[4], Olga Kazakova[2], Joerg Wunderlich[3,5], and Hans Werner Schumacher[1]

**Affiliations:**
1) Physikalisch-Technische Bundesanstalt, Bundesalle 100, D-38116 Braunschweig
2) National Physical Laboratory, Teddington, TW11 0LW, United Kingdom
3) Institute of Physics, Academy of Science of the Czech Republic, Cukrovarnická 10, 162 00 Praha 6, Czech Republic
4) Faculty of Mathematics and Physics, Charles University in Prague, Ke Karlovu 3, 121 16 Prague 2, Czech Republic
5) Hitachi Cambridge Laboratory, Cambridge CB3 0HE, United Kingdom
6) School of Physics and Astronomy, University of Nottingham, Nottingham NG7 2RD, UK
7) Thin Film Magnetism Group, Cavendish Laboratory, University of Cambridge, JJ Thomson Avenue, Cambridge CB3 0HE, UK

**Abstract**

In magnetic nanowires with perpendicular magnetic anisotropy (PMA) magnetic domain walls (DW) are narrow and can move rapidly driven by current induced torques. This enables important applications like high-density memories for which the precise detection of the position and motion of a propagating DW is of utmost interest. Today's DW detection tools are often limited in resolution, or acquisition speed, or can only be applied on specific materials. Here, we show that the anomalous Nernst effect provides a simple and powerful tool to precisely track the position and motion of a single DW propagating in a PMA nanowire. We detect field and current driven DW propagation in both metallic heterostructures and dilute magnetic semiconductors over a broad temperature range. The demonstrated spatial resolution below 20 nm is comparable to the DW width in typical metallic PMA systems.

# Article

Recent concepts for high-density memory, logic, and sensor devices [1] rely on the controlled positioning and propagation of narrow magnetic domain walls (DW) [2] in nanowires with perpendicular magnetic anisotropy (PMA) [3,4,5,6,7]. To study and develop such systems requires a reliable high-resolution tool for detecting the DW position inside the wire. While magneto-optical microscopy is limited in spatial resolution [3,8], high-resolution imaging such as spin resolved electron microscopy [9], nanomagnetometry [10], or magnetic force microscopy (MFM) [11] can be complex, time consuming and, for MFM, invasive. The anomalous Hall effect (AHE, Fig. 1(a)) [12] allows to probe the position and motion of a DW inside a PMA Hall cross with nanometer resolution [13,14], however the DW position inside the PMA wire itself is not accessible. Other electrical measurements like giant magneto resistance detection can only be applied on specific spin valve nanowires [15,16]. Over the last years, the field of spin-caloritonics [17] has explored the interplay of heat and spin currents in spintronic materials and devices. With respect to DW devices thermal spin transfer torque [18,19], thermally driven DW motion [20,21,22], and the magneto Seebeck contribution of an individual DW [23] were studied. At the same time the anomalous Nernst effect (ANE) moved into focus from being part of careful analysis of spin-caloritronic measurements [24,25,26] to detection of magnetization reversal [27,28] and magnetization dynamics [29,30] in magnetic thin films.

Here, we show that ANE provides a powerful tool to detect DW propagation in magnetic nanowires with nano scale resolution. Using a simple thermoelectrical measurement setup we probe current induced, magnetic field induced, and MFM induced DW propagation as well as DW depinning from individual nano scale pinning sites. To highlight the generic character of this method, we apply it on two distinct ferromagnetic PMA systems, namely: metallic Pt/CoFeB/Pt wires with high Curie temperature $T_C$, and wires patterned from a (GaMn)(AsP) magnetic semiconductor film [8] with $T_C$ below room temperature. We demonstrate DW position detection with spatial resolution down to 20 nm comparable to the DW width in typical PMA materials with potential for further improvement.

As sketched in Fig. 1(a) AHE has been used to detect DW propagation within PMA Hall crosses [13]. The AHE voltage $V_{AHE}$ is proportional to the average out-of-plane magnetization in the cross area which depends on the DW position. However an AHE signal is only generated in the cross region where the probe current $I$ is applied. By instead using *thermo*electrical measurements of the ANE (in analogue to the *electrical* AHE) the sensitive region can be extended over the whole wire length as shown in Fig. 1(b). To do so, a transverse in-plane thermal gradient $\nabla T_y$ is created by a heater line parallel to the wire. $\nabla T_y$ is perpendicular to the PMA magnetization $\mathbf{M}^{\downarrow,\uparrow}$ thus resulting in an ANE voltage $V_{ANE}$ between the wire ends. For fully saturated magnetization of the complete wire the ANE voltage is given by

$$V_{ANE}^{max} = -N_{ANE}\mu_0 m_z l \overline{\nabla T_y} \quad (1).$$

Here $N_{ANE}$ is the ANE coefficient per magnetic moment, $\mu_0$ is the vacuum permeability, $m_z$ is the *z*-component of the magnetization (with $m_z = \pm M_S$ the saturation magnetization), $l$ is the wire length and $\overline{\nabla T_y}$ is the average of $\nabla T_y$ over the wire length. A single DW at position $x_{DW}$ inside the wire divides the magnetisation distribution into two domains with opposite perpendicular magnetization $m_z^{\downarrow} = -m_z^{\uparrow}$. If the DW is positioned in the centre of the wire (defined $x_{DW} = 0$) the ANE contributions of the two domains compensate and $V_{ANE}(x_{DW} = 0)$ vanishes. DW motion changes the ratio between the two ANE contributions and hence $V_{ANE}$. For a constant $\nabla T_y$ over the wire length $V_{ANE}(x_{DW})$ depends linearly on $x_{DW}$, thus allowing direct thermoelectrical detection of the DW position with nanoscale resolution using:

$$x_{DW} = -\frac{V_{ANE}(x_{DW})}{2N_{ANE}\mu_0 M_S \nabla T_y} \quad (2)$$

Note that the assumption of constant $\nabla T_y(x)$ is only valid for constant wire width and for samples with heater line longer than the wire. A more general expression of $V_{ANE}(x_{DW})$ for spatially varying $\nabla T_y(x)$ is derived in the supplementary information.

Fig. 1(c) shows a false colour electron microscope image of a Pt/CoFeB(0.6 nm)/Pt nanowire (yellow) with an adjacent Pt heater line. Fig. 1(d) shows the simulated temperature distribution upon application of $P_{heat}$ = 5.1 mW to the heater line. For details on simulations and temperature calibrations, see supplementary information. From the temperature profile in Fig. 1(e) taken along the dashed line in 1(d) we estimate a temperature drop of 45 mK across the wire. Furthermore the simulation yields $\overline{\nabla T_y}$ = 67 ± 7 mK/µm over the whole wire length. Fig. 1(f) shows typical ANE data of magnetization reversal in out-of-plane fields for Pt/CoFeB/Pt (black) and (GaMn)(AsP) (red) devices. The metallic wire is measured at room temperature, whereas the semiconducting wire (Fig 3(a)) is measured at $T$ = 65.5 K, below $T_C$. Both wires show a clear hysteretic behaviour in $V_{ANE}$ with the square loop indicating magnetisation reversal by fast DW motion. For both samples, $V_{ANE}$ scales linearly with the applied heater power and hence with $\nabla T_y$, as expected for a thermoelectrical signal (not shown). For the given heater parameters we obtain values of $V_{ANE}^{max}$ of 350 nV for CoFeB and of 2 µV for (GaMn)(AsP). Using (1) with experimental values of $\mu_0 M_S$ of 1.3 T (CoFeB) and of 22.9 mT ((GaMn)(AsP)) we estimate the ANE coefficients of $N_{ANE}$ = of -0.37 µV(KT)$^{-1}$ (CoFeB) and of -1,49 µV(KT)$^{-1}$ ((GaMn)(AsP)) at the given temperatures.

We will now focus on ANE-based DW detection in CoFeB devices at room temperature. Fig 1(g) shows an ANE reversal curve of a similar device as characterized in 1(f). As shown in 1(c) the wire contains a notch in the centre (at x = 0). It can act as a pinning site for a propagating DW provided that the depinning field threshold is higher than the field threshold for nucleating a reversed domain at the end of the wire. The MFM image in the inset shows a DW pinned at the notch of the same device. For the reversal curve the sample was first saturated by an applied negative field of $B_{app}$ = -200 mT. Then the field was swept to positive fields and $V_{ANE}$ was recorded. For negative saturation, $V_{ANE}$ remains constant at about -175 nV. At a field of ~ 25 mT, $V_{ANE}$ sharply rises to a nearly zero value. Here, a reversed domain has first nucleated at the narrow end of the nanowire [31] and then propagated to the notch where it remains pinned. As the notch is centred between the two contacts $V_{ANE}(x_{DW} = 0)$ vanishes as discussed above. Around 30 mT, the DW depins from the notch and continues propagation through the remaining part of the wire, completing the magnetization reversal ($V_{ANE}$ = +175 nV). The steep transitions indicate rapid DW propagation, with a propagation field threshold well below the applied field.

For more detailed investigations of high resolution DW detection, we use the local stray field $B_{tip}$ of an MFM tip to control the DW position in CoFeB devices on the nanometer scale. As sketched in Fig 2(a) a high moment MFM probe (Methods) is scanned across the wire (as schematically indicated by the red line). During the scan, $B_{tip}$ nucleates a reversed domain and displaces the DW along the wire thereby changing the ANE response as shown in Fig. 2(b). Here, $V_{ANE}$ is plotted as function of tip position $x_{tip}$ for $B_{app}$ = 0 mT. The right scale shows $x_{DW}$ derived from $V_{ANE}$. The wire geometry relative to $x_{tip}$ is indicated by the top image. Before the scan, the sample is saturated in negative field $B_{app}$ = -100 mT and $V_{ANE} \cong$ -330 nV. The scan starts outside the first electrical contact ($x_{tip} \leq$ -3.2 µm) and thus outside the sensitive region of ANE measurements and $V_{ANE}$ remains constant. Once the probe has crossed the first contact ($x_{tip} \cong$ -3.2 µm), an increase of $V_{ANE}$ is observed. The small plateaus separated by sharp transitions indicate stepwise DW motion from one local energy minimum (pinning site) to the next. The energy minima are due to local fluctuations of the magnetic properties, *e.g.* from edge roughness, local defects or local thickness variation of the film. The widest plateau, and hence the strongest DW pinning, is observed at the notch (around $x_{tip}$ = 0, $V_{ANE}$ = 0). Upon entering the tapered region, the DW can reduce its length, and thus self-energy, by moving ahead of the MFM tip towards the narrowest part of the notch. After depinning from the notch, DW propagation again reveals several pinning and depinning events indicated by small plateaus before the DW exits the measurement region at the second contact.

Fig. 2(c,d) show zooms of two pairs of plateaus marked by the coloured arrows in (b). $x_{DW}$ is plotted vs. $x_{tip}$. Statistical analysis of the two plateaus in 2(c) (red and black arrows in (b)) yields absolute DW positions of $x_{DW}$ = (1667 ± 18) nm (black) and $x_{DW}$ = (1816 ± 19) nm (black). The low positioning uncertainty below 20 nm clearly demonstrates nano scale spatial resolution of ANE DW detection. Estimating the DW width by $w_{DW} = \pi\sqrt{AK_{eff}^{-1}}$ [32] with $A$ = 2.5·10$^{-11}$ Jm$^{-1}$ the exchange stiffness and $K_{eff}$ = 5·10$^5$ Jm$^{-3}$ the effective anisotropy yields $w_{DW}$ ≅ 20 nm. The obtained spatial resolution thus matches the most relevant physical length scale in our system. 2(d) shows the two closest plateaus of the measurement that are clearly separated within the noise level (blue and orange arrows). The data yields a spatial separation of the two pinning sites of only (64 ± 30) nm demonstrating the capability of detecting depinning processes from individual pinning sites on the nanometer scale. Furthermore it shows the potential of MFM nano manipulation of a DW in a magnetic wire.

In Fig. 2(e) we investigate the effect of nonzero $B_{app}$ on tip induced DW motion. Here, three curves at $B_{app}$ = -4.0 (blue), 3.7 (red) and 8.3 mT (black) are shown. Note that the curves were taken at different experimental parameters (higher MFM tip velocity, lower $P_{heat}$) leading to a different signal-to-noise ratio than in 2(b)-(d). The effect of $B_{app}$ is clearly observed. $B_{app}$ = -4 mT is antiparallel to the local stray field underneath the tip ($B_{tip} > 0$). Still the remaining total field under the MFM tip $B_{tot} = B_{tip} + B_{app}$ is sufficient to nucleate and propagate a reversed DW leading to a basically linear displacement of $x_{DW}$ with $x_{tip}$ outside the notch region. Contrarily, if the probe field coincides with the direction of $B_{app}$ (black, red), the total field increases and the DW propagates ahead of the tip, resulting in steeper curves and less pronounced pinning inside the wire. For $B_{app}$ = -8.3 mT, (black) pinning inside the wire becomes negligible and DW propagation is only hindered by pinning at the notch.

With respect to Fig. 3, we now discuss thermo-electrical DW detection of field and current driven DW propagation in a 2 µm wide and 50 um long (GaMn)(AsP)wire. Fig. 3(a) shows the device in false colours. It contains a freestanding nucleation strip line (blue) crossing the left contact area of the magnetic wire (yellow). Application of a 10 ms long nucleation pulse of 20 mA to the strip line generates an Oersted field which forms a reversed magnetized domain with a single DW present on the left hand side of the magnetic wire.

Fig 3 (b) shows the variation of the normalized ANE signal ($V_{ANE}$ normalized by $V_{ANE}^{max}$) when the DW creeps along the magnetic wire driven by a small magnetic field. First, the wire is saturated in a negative saturation field of -20 mT before the field is swept at 0.25 mT/s from 0 to +0.5 mT. The DW is nucleated at $B_{app} \cong 0.05$ mT (blue arrow) and the subsequent wide $V_{ANE}$ plateau indicates DW pinning at a non-intentional pinning site until $B_{app}$ exceeds 0.2 mT. Note that the peak at the beginning of the plateau is induced by electrical cross talk from the nucleation pulse. With increasing field, additional plateaus are observed which indicate further DW pinning during the DW propagation along the bar. The normalized ANE signal of plateaus (i - iii) correspond to DW positions at $x_{DW}$ = -20 um, +4 µm and + 15 µm. This pinning scenario is confirmed by magneto-optical Kerr effect (MOKE) micrographs of the pinned DW states shown in the insets (i-iii) of Fig. 3(b).

Fig. 3(c) shows ANE detection of current pulse driven DW propagation at $B_{app}$ = 0 mT [8]. In our experiments, 1us long pulses of current densities ranging from j = 6.3 - 10.7×10$^9$ A/m$^2$ are applied. Before each series of pulses of constant current amplitude, the wire magnetization is saturated and a single DW is nucleated. The first $V_{AHE}$ data-point is measured directly after nucleation and subsequent data-points are taken after each individual current pulse. At positive current, the ANE signal changes from negative to positive saturation indicating DW propagation along the wire from left to right. In contrast spin torque current pulses of opposite polarity move the DW backwards (supplementary information). At high current densities $j$ > 10 ×10$^9$ A/m$^2$ only few pulses are sufficient to propagate the DW through the entire wire. From the measured change of $x_{DW}$ during a 1 μs pulse one can directly derive the DW velocity up to 30 m/s. In contrast, at low current densities, $j < 9 \times 10^9$ A/m$^2$, many consecutive pulses are required to propagate the DW along the entire wire. Note that at lower current densities spin torque driven DW propagation is again affected by pinning. $V_{ANE}$ thus shows pinning plateaus and does not increase linearly with the number of pulses. This again underlines the importance of high resolution tools for studies of current driven DW propagation as provided by ANE detection. Only with easily accessible high resolution tools at hand local pinning at defects can be pinpointed and a detailed understanding of the DW propagation mechanism is possible

As shown in Fig 2(b) we have already demonstrated a spatial resolution of ANE detection below 20 nm which is comparable to the DW width in our metallic PMA system. This resolution is also comparable to typical resolutions obtained by high-resolution imaging techniques such as ambient MFM. Note that in contrast to MFM or magneto transport detection our ANE detection method can be considered as non-invasive as no local fields or probe currents are applied to the magnetic wire. Better signal-to-noise-ratios and thus higher resolution of ANE detection should be obtainable by optimization of the sample layout. Decreasing the wire-heater separation could allow to increase $\nabla T_y$ and hence the spatial resolution by about 25% with only a moderate increase of the overall wire temperature. Further improvement of the sample design, e.g. by using membrane substrates, could significantly increase $\nabla T_y$, and could enable spatial resolutions down to the few nm range. Furthermore, high bandwidth data acquisition might in the future enable studies of fast DW dynamics. Thus, ANE provides a powerful and versatile tool for future studies of DW propagation dynamics induced by fields, spin-torques or spin-orbit-torques in a broad variety of spintronic materials.

**Methods:**

**Samples:** CoFeB samples were deposited using magnetron sputtering and consist of Ta (4 nm) / Pt (3 nm) / Co$_{0.6}$Fe$_{0.2}$B$_{0.2}$ ($t_{CoFeB}$) /Pt (3 nm) multilayers (with $t_{CoFeB}$ = 0.6 - 1.3 nm) grown at 8 x10$^{-3}$ mbar Ar pressure in a system with a base pressure of 8x10$^{-8}$ mbar. The substrate was 300 nm thermal SiO$_2$ on p-doped Si. The magnetic thin films have a large perpendicular anisotropy and show sharp coercive switching. The multilayer films were patterned into nanowires by electron beam lithography and ion beam etching. Electrical contacts to the nanowires, heater lines and thermometer lines were defined in a lift-off process using electron beam lithography and sputter deposition of a 5 nm Ta adhesion layer and 30 nm Pt.

(GaMn)(AsP) samples comprise a 2 µm wide and 50 µm long wire patterned by electron-beam lithography along the [110] crystal axis of the Ga$_{0.94}$Mn$_{0.06}$As$_{0.91}$P$_{0.09}$ epilayer grown by low-temperature molecular beam epitaxy on GaAs substrate and buffer layers [33]. The Curie temperature of the annealed ferromagnetic semiconductor material was 115 K and the conductivity was 230 Ω$^{-1}$cm$^{-1}$. Two 200 nm wide Pt heater wires were patterned 3 µm above and below the central (GaMn)(AsP)bar. After structuring the magnetic wire and heater lines, a 180 nm thick PMMA layer was spun on top of the structure and high-dose e-beam irradiation was used to cross-link two 20 µm wide areas, each intersecting the microbar on one of its ends. Subsequently, a 300 nm thick Cr/Au strip-line was deposited by thermal evaporation on top of each cross-linked PMMA areas and followed by the lift-off procedure to generate nucleation strip-lines at both ends of the bar electrically insulated from the magnetic wire.

**ANE Measurements:** In-plane thermal gradients in the *y*-direction were generated by electrical heater lines parallel to the nanowires. For CoFeB devices, an oscillating heat gradient was generated by application of AC heater powers up to about 7.4 mW with AC frequency $f_{heat}$ = 17 Hz. $V_{ANE}$ was measured using lock-in detection at the second harmonic. For (GaMn)(AsP) devices, the positive, and negative half waves of an alternating current were flowing along upper and lower heater lines, respectively, phase shifted by $\pi$. They generate an alternating transverse temperature gradient oscillating at the frequency of the alternating current. The ANE signal is therefore measured at the fundamental frequency $f_{heat}$ using a login amplifier (supplementary information). Adjusting the cooling power of the cryostat, the time-averaged temperature of the central magnetic wire was kept constant at $T$ = 65.5 K for all applied heater currents of 1 - 7 mA amplitude.

**MFM induced DW manipulation:** MFM measurements were conducted using an NT-MDT Aura SPM system in ambient atmosphere. Scanning was performed in semi-contact (tapping) mode, Nanosensors PPP-MFMR probes were used. The sample was bonded for electrical measurements before being installed on top of the system's out-of-plane electromagnet. The output voltages from the lock-in amplifier were recorded using the external inputs on the SPM system. Measurements involving MFM probe induced remagnetisation and DW motion were done by conducting a topography image of the device area (with slow scan axis along x), while simultaneously recording the Nernst signal from the device at each point of the scan.

**Magneto optical Kerr effect (MOKE) imaging:** For MOKE imaging of pinned DW states in (GaMn)(AsP)samples the following procedure was taken. First, the field was increased until an increase of $V_{ANE}$ due to DW propagation to a given pinning site was detected. Then, the magnetic field was reduced to zero for MOKE imaging of the DW position in the remanent state. The domain contrast was achieved by subtraction of the image acquired during this procedure from a reference image taken at saturation.

**Acknowledgements:** This work was cofunded by EU and EMRP within JRP EXL04 SpinCal. H.W.S. acknowledges funding by DFG SPP 1538 SpinCaT. J.W. acknowledges funding from the European Research Council under the European Union's Seventh Framework Programme (FP7/2007-2013) / ERC grant agreement n 610115. P.N acknowledges funding by the Grant Agency of the Czech Republic under Grant No. 14-37427G and by the Grant

Agency of Charles University Grant no. 1910214. O.K. acknowledges support by the UK government's Department for Business, Energy and Industrial Strategy. We thank André Müller for preparation of the 3D graphics.

**Author contributions:**

P.K. and J.We. performed and analyzed ANE experiments on CoFeB samples. A.F.S, V.S., T.J., P.N., and J.Wu. carried out the electrical and optical measurements on the (GaMn)(AsP) devices. Z.S and J.Wu. designed and patterned the (GaMn)(AsP) device. R.P.Ca. prepared the magnetic semiconductor film. P.K. designed and patterned the CoFeB devices. R.M. and J.-H.L prepared the CoFeB films. X.H. performed thermal simulations and MFM measurements. R.P.Co., O.K., J.Wu., and H.W.S. supervised the research. All authors contributed to preparation of the manuscript.

**References:**

[1]   C. Chappert, A. Fert, & F.N.V. Dau: *The emergence of spin electronics in data storage*, Nature Mater. **6**, 813 (2007).
[2]   P. J. Metaxas, J. P. Jamet, A. Mougin, M. Cormier, J. Ferré, V. Baltz, B. Rodmacq, B. Dieny, and R. L. Stamps: *Creep and Flow Regimes of Magnetic Domain-Wall Motion in Ultrathin Pt/Co/Pt Films with Perpendicular Anisotropy*, Phys. Rev. Lett. **99**, 217208 (2007).
[3]   Kwang-Su Ryu, Luc Thomas, See-Hun Yang, and Stuart S. P. Parkin: *Current Induced Tilting of Domain Walls in High Velocity Motion along Perpendicularly Magnetized Micron-Sized Co/Ni/Co Racetracks*, Applied Physics Express **5**, 093006 (2012).
[4]   See-Hun Yang, Kwang-Su Ryu & Stuart Parkin: *Domain-wall velocities of up to 750 m s−1 driven by exchange-coupling torque in synthetic antiferromagnets,* Nature Nanotechnology **10**, 221 (2015).
[5]   S. Dutta Gupta, S. Fukami, C. Zhang, H. Sato, M. Yamanouchi, F. Matsukura, and H. Ohno: *Adiabatic spin-transfer-torque-induced domain wall creep in a magnetic metal*, Nature Physics, **12**, 333–336 (2016).
[6]   S.Emori, U.Bauer, S.-M.Ahn, E. Martinez, G. S. D.Beach *Current-driven dynamics of chiral ferromagnetic domain walls*, Nature Mater. 12, 611 (2013).
[7]   Yoko Yoshimura, Kab-Jin Kim, Takuya Taniguchi, Takayuki Tono, Kohei Ueda, Ryo Hiramatsu, Takahiro Moriyama, Keisuke Yamada, Yoshinobu Nakatani & Teruo Ono: *Soliton-like magnetic domain wall motion induced by the interfacial Dzyaloshinskii–Moriya interaction*, Nature Physics **12**, 157 (2016).
[8]   E. De Ranieri, P. E. Roy, D. Fang, E. K. Vehsthedt, A. C. Irvine, D. Heiss, A. Casiraghi, R. P. Campion, B. L. Gallagher, T. Jungwirth, and J.Wunderlich: *Piezoelectric control of the mobility of a domain wall driven by adiabatic and non-adiabatic torques*, Nature Materials **12**, 808 (2013).
[9]   H.P.Oepen, H.Hopster, SEMPA studies of thin film structures and exchange coupled layers, p.137-167 in the book "Magnetic Microscopy of Nanostructures" eds H.Hopster, H.P.Oepen, Springer, 2003.
[10]  J.-P. Tetienne,T. Hingant, J.-V. Kim, L. Herrera Diez, J.-P. Adam, K. Garcia, J.-F. Roch, S. Rohart, A. Thiaville, D. Ravelosona, V. Jacques: *Nanoscale imaging and control of domain-wall hopping with a nitrogen-vacancy center microscope*, Science **344**, 1366 (2014).
[11]  A. Yamaguchi, T. Ono, and S. Nasu, K. Miyake, K. Mibu, T. Shinjo: *Real-space observation of current-driven domain wall motion in submicron magnetic wires*. Phys. Rev. Lett. 92, 077205 (2004).
[12]  Naoto Nagaosa, Jairo Sinova, Shigeki Onoda, A. H. MacDonald, N. P. Ong: *Anomalous Hall Effekct*. Rev. Mod. Phys. **82**, 1539 (2010).
[13]  J. Wunderlich, D. Ravelosona, C. Chappert, F. Cayssol, V. Mathet, J. Ferre, J. P. Jamet, and A. Thiaville, *Influence of geometry on domain wall propagation in a mesoscopic wire,* IEEE Trans. Mag. **37**, 2104 (2001).
[14]  T. Koyama, D. Chiba, K. Ueda, K. Kondou, H. Tanigawa, S. Fukami, T. Suzuki, N. Ohshima, N. Ishiwata, Y. Nakatani, K. Kobayashi & T. Ono: "*Observation of the intrinsic pinning of a magnetic domain wall in a ferromagnetic nanowire*" Nature Materials 10, 194–197 (2011).
[15]  C. Burrowes, A. P. Mihai, D. Ravelosona, J.-V. Kim, C. Chappert, L. Vila, A. Marty, Y. Samson, F. Garcia-Sanchez, L. D. Buda-Prejbeanu, I. Tudosa, E. E. Fullerton and J.-P. Attané: *Non-adiabatic spin-torques in narrow magnetic domain walls* Nature Physics DOI: 10.1038/NPHYS1436 (2010).
[16]  T. Ono, H. Miyajima, K. Shigeto, K. Mibu, N. Hosoito, T. Shinjo: *Propagation of a magnetic domain wall in a submicrometer magnetic wire* Science **284**, 468–470 (1999).
[17]  G.E.W. Bauer, E. Saitoh, and B. J. van Wees: *Spin caloritronics*, Nature Materials **11**, 391 (2012).
[18]  Alexey A. Kovalev and Yaroslav Tserkovnyak: *Thermoelectric spin transfer in textured magnets,* Phys. Rev. B 80, 100408(R) (2009).
[19]  P. Yan, X. S. Wang, and X. R. Wang, *All-Magnonic Spin-Transfer Torque and Domain Wall Propagation*, Phys. Rev. Lett. **107**, 177207 (2011)
[20]  D. Hinzke and U. Nowak: *Domain Wall Motion by the Magnonic Spin Seebeck Effect*, Phys. Rev. Lett **107**, 027205 (2011)
[21]  J. Torrejon, G. Malinowski, M. Pelloux, R. Weil, A. Thiaville, J. Curiale, D. Lacour, F. Montaigne, and M. Hehn: *Unidirectional Thermal Effects in Current-Induced Domain Wall Motion*, Phys. Rev. Lett. **109**, 106601 (2012).

[22] W. Jiang, P. Upadhyaya, Y. Fan, J. Zhao, M. Wang, L.-T. Chang, M. Lang, K.L. Wong, M. Lewis, Y.-T. Lin, J. Tang, S. Cherepov, X. Zhou, Y. Tserkovnyak, R. N. Schwartz, and K. L. Wang: *Direct Imaging of Thermally Driven Domain Wall Motion in Magnetic Insulators*, Phys. Rev. Lett. **110**, 177202 (2013).

[23] P. Krzysteczko, X. Hu, N. Liebing, S. Sievers, and H.W. Schumacher: *Domain wall magneto Seebeck effect*, Phys. Rev. **92**, 140405 (2015).

[24] S. Y. Huang, W. G. Wang, S. F. Lee, J. Kwo, and C. L. Chien: *Intrinsic Spin-Dependent Thermal Transport,* Phys. Rev. Lett. 107, 21660 (2011).

[25] T. Kikkawa, K. Uchida, Y. Shiomi, Z. Qiu, D. Hou, D. Tian, H. Nakayama, X.-F. Jin, and E. Saitoh: *Longitudinal Spin Seebeck Effect Free from the Proximity Nernst Effect,* Phys. Rev. Lett. **110**, 067207 (2013).

[26] D. Meier, D. Reinhardt, M. Schmid, C. H. Back, J.-M. Schmalhorst, T. Kuschel, and G. Reiss: *Influence of heat flow directions on Nernst effects in Py/Pt bilayers*, Phys. Rev. B. 88, 184425 (2013).

[27] Masaki Mizuguchi, Satoko Ohata, Ken-ichi Uchida, Eiji Saitoh, and Koki Takanashi: *Anomalous Nernst Effect in an $L_{10}$-Ordered Epitaxial FePt Thin Film,* Appl. Phys. Express **5** 093002 (2012)

[28] K. Baumgaertl, F. Heimbach, S. Maendl, D. Rueffer, A. Fontcuberta i Morral, and D. Grundler: *Magnetization reversal in individual Py and CoFeB nanotubes locally probed via anisotropic magnetoresistance and anomalous Nernst effect* Appl. Phys. Lett. **108**, 132408 (2016); doi: 10.1063/1.4945331

[29] J.M. Bartell, D.H. Ngai, Z. Leng, and G.D. Fuchs: *Towards a table-top microscope for nanoscale magnetic imaging using picosecond thermal gradients,* Nat. Comm. **6**, 8460, DOI: 10.1038/ncomms9460.

[30] H. Schultheiss, J. E. Pearson, S.D. Bader, and A. Hoffmann: *Thermoelectric Detection of Spin Waves,* Phys. Rev. Lett. **109**, 237204 (2012).

[31] R. Mansell, A. Beguivin, D.C.M.C. Petit, A. Fernández-Pacheco, J.H. Lee, and R.P. Cowburn: *Controlling nucleation in perpendicularly magnetized nanowires through in-plane shape*, Appl. Phys. Lett. **107**, 092405 (2015); http://dx.doi.org/10.1063/1.4930152.

[32] J. M. D. Coey: *Magnetism and Magnetic Materials,* Cambridge University Press, New York (2009).

[33] K. Y. Wang, K. W. Edmonds, A. C. Irvine, G. Tatara, E. De Ranieri, J. Wunderlich, K. Olejnik, A. W. Rushforth, R. P. Campion, D. A. Williams, C. T. Foxon, and B. L. Gallagher: *Current-driven domain wall motion across a wide temperature range in a (Ga,Mn)(As,P) device*, Appl. Phys. Lett. **97**, 262102 (2010).

**Figures:**

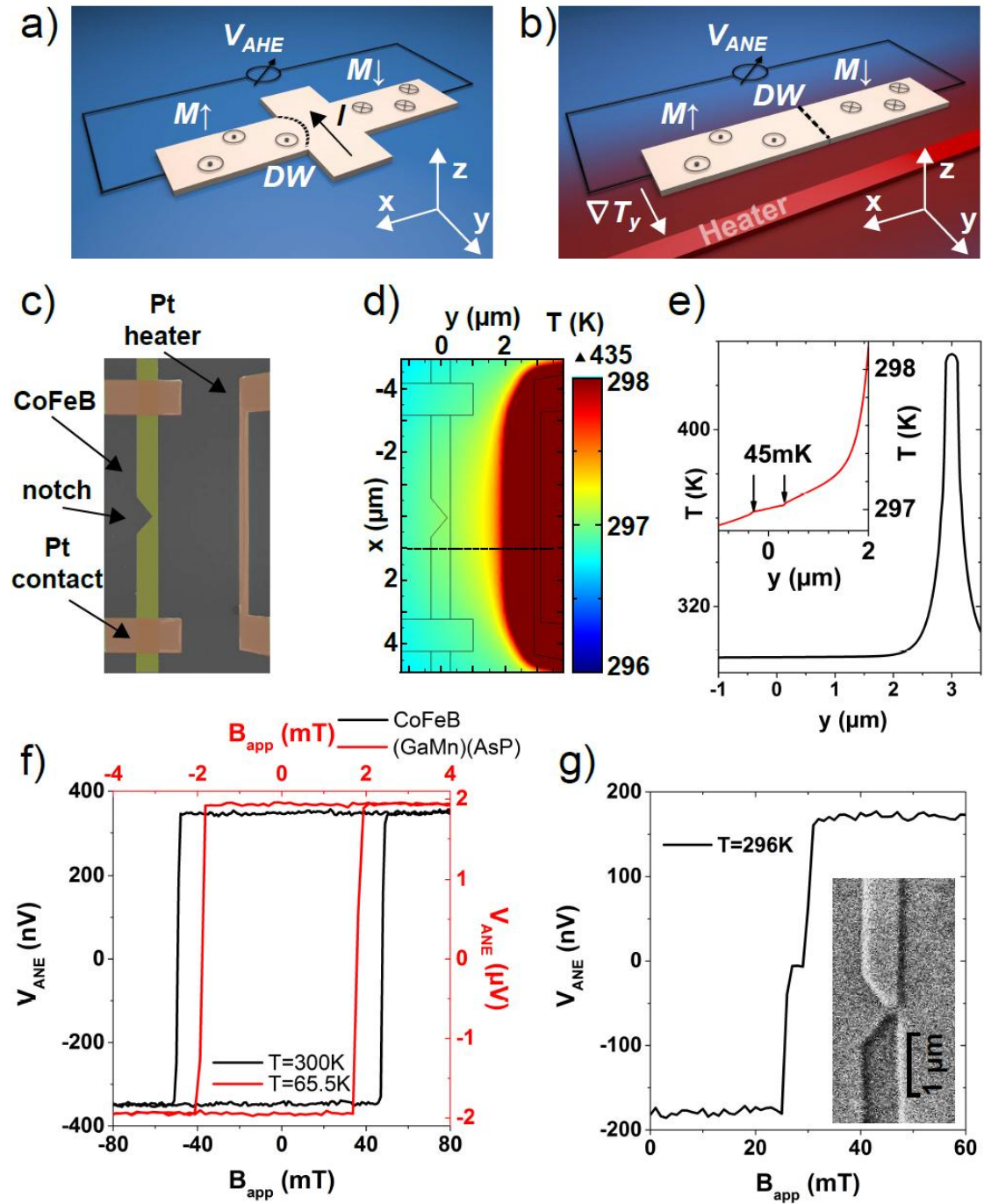

**Figure 1:** Principle of ANE based DW detection: (a) AHE DW detection by probe current $I$ inside a Hall cross. (b) ANE DW detection in a PMA wire: a transverse temperature gradient $\nabla T_y$ generates an ANE voltage $V_{ANE}$ along the wire depending on $x_{DW}$. (c) False colour electron micrograph of CoFeB nanowire (yellow) with Pt contacts and heater line (red). The wire $l$ length between the contacts is 6.4 μm. The wire is 600 nm wide. A 200 nm wide notch is situated at the centre ($x = 0$). (d) Colour map of the simulated temperature distribution for electrical heating with $P_{heat}$ = 5.1 mW. (e) linear temperature profile in y-direction along the dashed line in (d). The heater locally heats up to 435 K. The main temperature drop occurs within only 1 µm lateral distance. Inset: zoom of the temperature profile. Wire boundaries are marked by arrows around $y$ = 0. The simulation yields a wire temperature of $T$ = 297 K and a temperature drop of 45 mK across the wire. (f) ANE reversal loops from full negative to positive saturation. $V_{ANE}$ is plotted vs. $B_{app}$. The black curve shows data of a CoFeB wire at room temperature. $P_{heat}$ = 7.4 mW. The red curve shows (GaMn)(AsP) data taken at $T$ = 65.5 K below $T_C$ = 115 K of the (GaMn)(AsP)film. $P_{heat}$ = 0.75 mW. (g) Pinning of a propagating DW at the notch of a CoFeB device. $V_{ANE}$ plotted as function of $B_{app}$ for a sweep from -200 mT to positive fields. $P_{heat}$ = 5.1 mW. Sweep rate 8.3 mT/s. The plateau at $V_{ANE}$ = 0 results from pinning of a propagating DW at the notch. Inset: MFM image of a pinned DW at the notch. The two contrasts correspond to opposite $m_z$ on the either side of the notch.

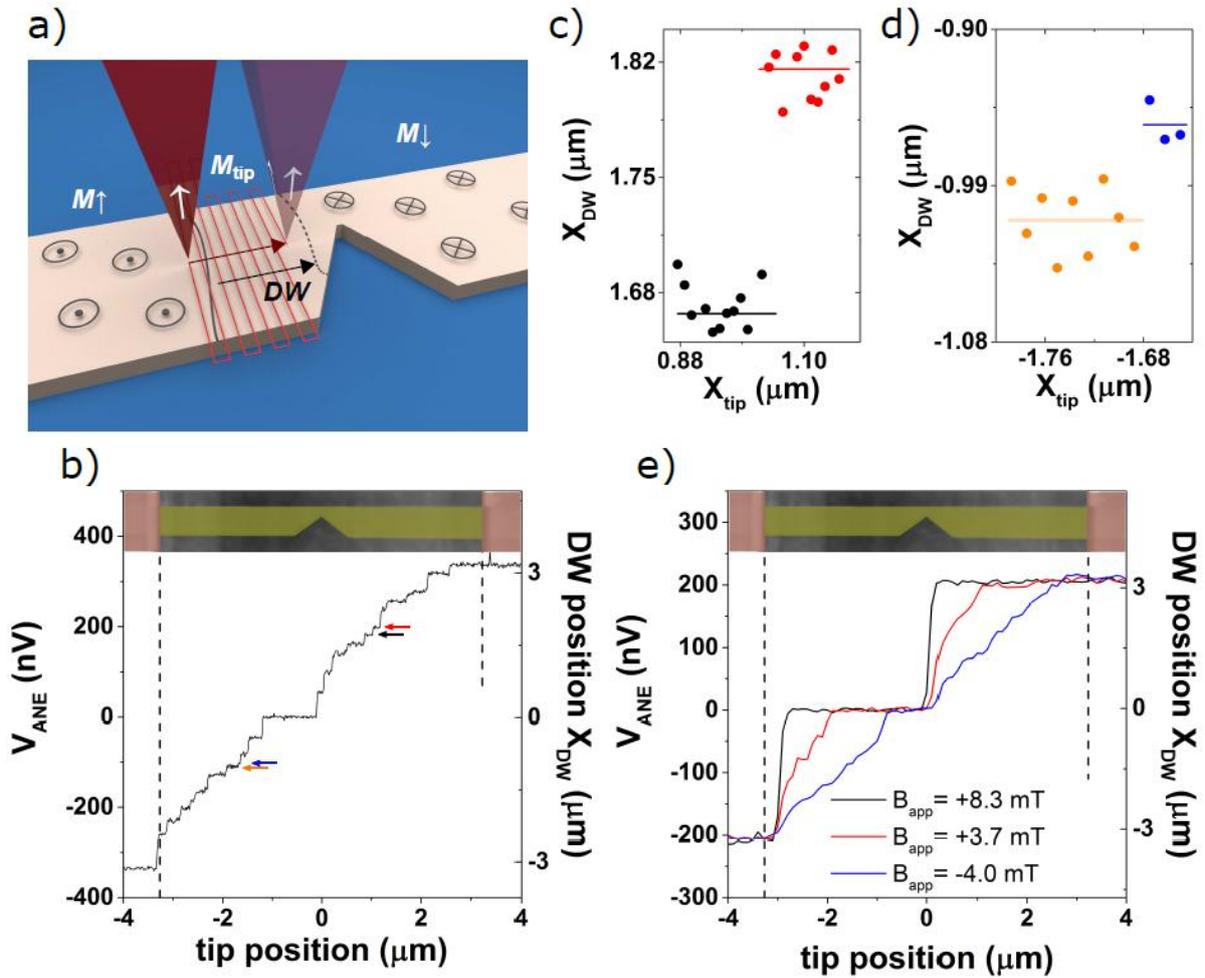

**Figure 2:** ANE detection of MFM controlled DW propagation in a CoFeB nanowire. (a) Scheme of MFM controlled DW propagation. An MFM tip is scanned across the wire. The stray field of the magnetic MFM tip nucleates a reversed domain and propagates the DW ahead of the tip. (b) ANE measurement of tip induced DW propagation at $B_{app} = 0$. $P_{heat} = 7.3$ mW. MFM tip *x*-velocity = 2.5 m/s. Top panel shows wire geometry relative to tip position. Plateaus indicate stepwise DW motion between pinning sites. The wide plateau near the centre corresponds to pinning at the notch. (c),(d) Zoom of ANE data of two plateau pairs indicated by the coloured arrows in (b). Data colours correspond the arrow colour in (d). (e) ANE measurement of tip induced DW propagation at applied fields of $B_{app}$ = -4.0 mT (blue), 3.7 mT (red), and 8.3 mT (black). $P_{heat}$ = 5.1 mW. Tip *x*-velocity = 30 nm/s. For higher fields parallel to the tip field the DW propagates ahead of the tip. For 8.3 mT pinning only occurs at the notch.

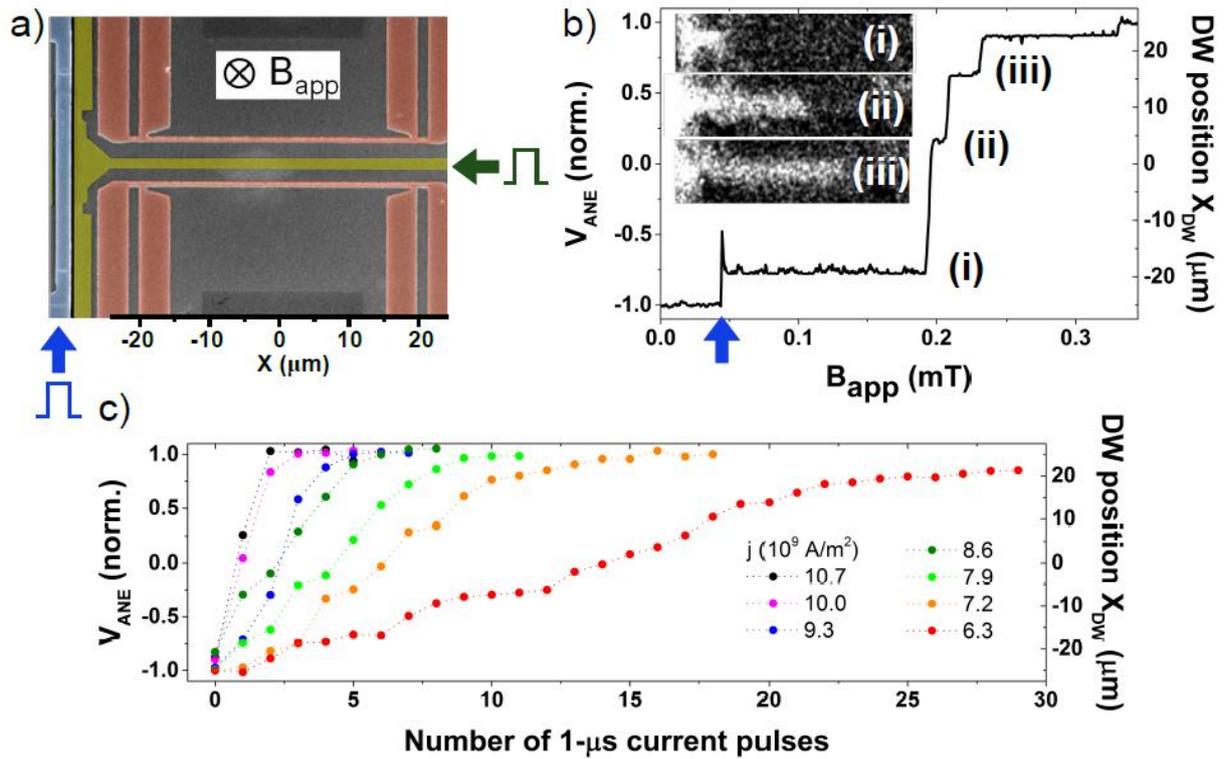

**Figure 3:** ANE DW detection in a (GaMn)(AsP) microwire. (a) False colour electron micrograph of the device. Yellow: wire, red: Pt heater line; blue: Au nucleation strip line. (b) ANE measurement during field induced DW propagation. Normalized $V_{ANE}$ vs. $B_{app}$. $P_{heat}$ = 0.5 mW. Sweep rate 0.25 mT/s. At $B_{app} \cong 0.05$ mT a DW is nucleated (blue arrow). Plateaus result from pinning at intrinsic unintentional pinning sites. Inset: MOKE microscope images corresponding to the three pinned DW states marked (i)-(iii). (c) ANE measurement of spin torque induced DW propagation. $B_{app} = 0$. ANE data is taken after application of 1μs pulses of current density $j$ as indicated. For high current density high DW velocities up to 30 m/s are found. For lower current density DW propagation is hindered by pinning at various non-intentional pinning sites.

**Supplementary information on**

# Nano scale thermo-electrical detection of magnetic domain wall propagation

*Patryk Krzysteczko, James Wells, Alexander Fernandez Scarioni, Zbynek Soban, Tomas Janda, Xiukun Hu, Vit Saidl, Richard P. Campion, Rhodri Mansell, Ji-Hyun Lee, Russell Cowburn, Petr Nemec, Olga Kazakova, Joerg Wunderlich, and Hans Werner Schumacher*

The supplementary information contains (**I**) the derivation of a generalized expression for dependence of the ANE voltage on the DW position in spatially varying thermal gradients, (**II**) details of ANE measurements on (GaMn)(AsP) devices, and (**III**) details on finite element simulations of temperature flux using the COMSOL package and on the corresponding thermal calibrations.

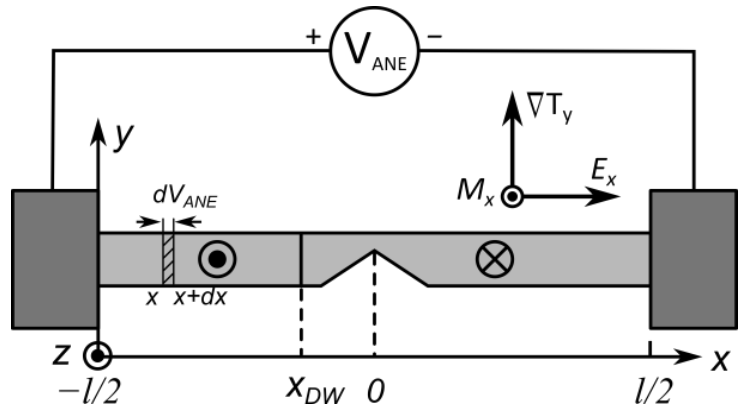

*Figure S1: Calculation of the ANE voltage between two contacts.*

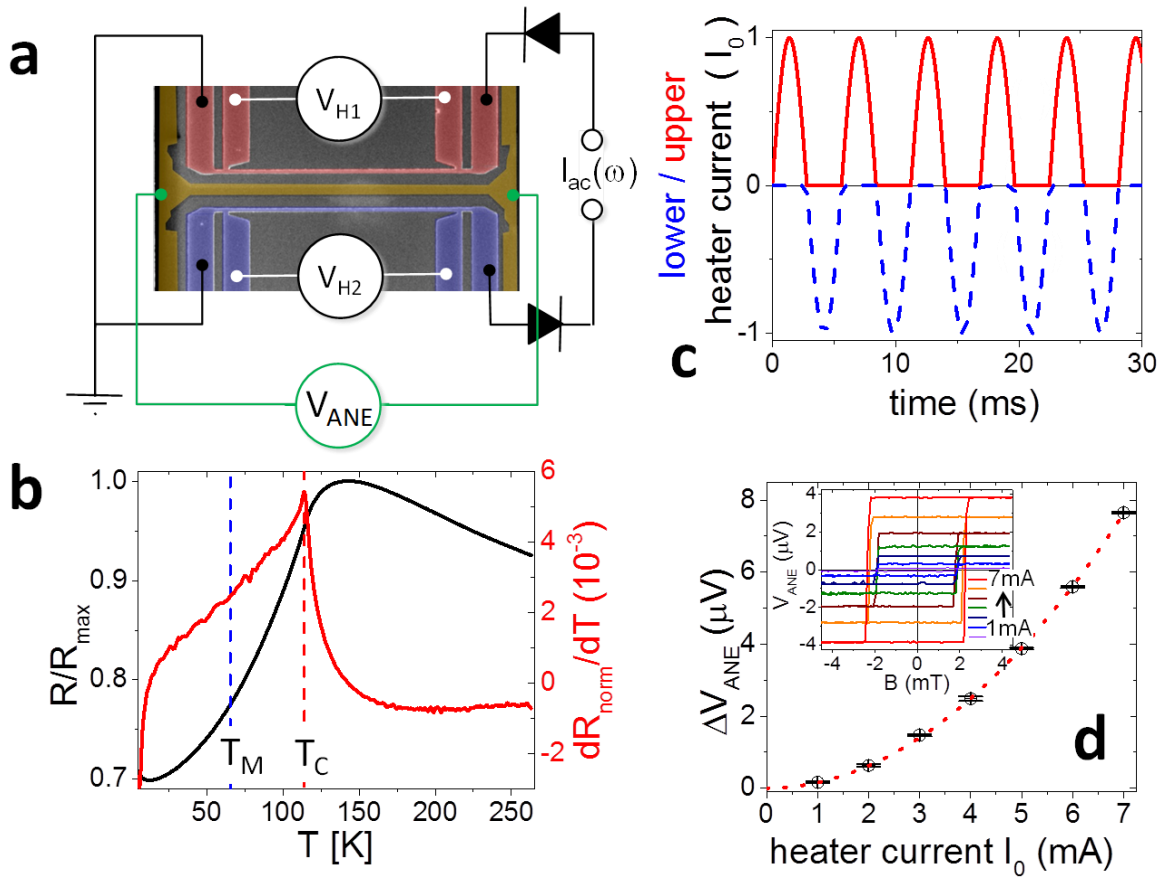

*Figure S2: (a) Illustration of ANE signal measurement; (b): Temperature dependence of the normalized resistance of the magnetic wire (black) and dR/dT identifying a Curie temperature of $T_C$ = 115 K (red); (c): Time dependence of the heater currents for upper (red) and lower (blue) bars. (d): Change of ANE signal during a complete magnetisation reversal as a function of heater current $I_0$. The inset shows the ANE reversal loops measured for heater currents of $I_0$ = 1, 2 , .. 7mA during a complete magnetisation reversal loop.*

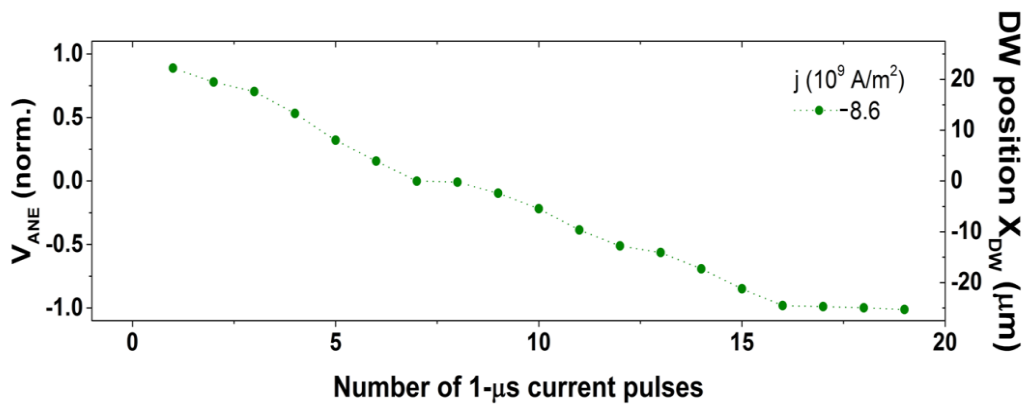

**Figure S3**: Spin torque induced backward propagation of a DW in the (GaMn)(AsP) wire by current pulses of negative polarity.

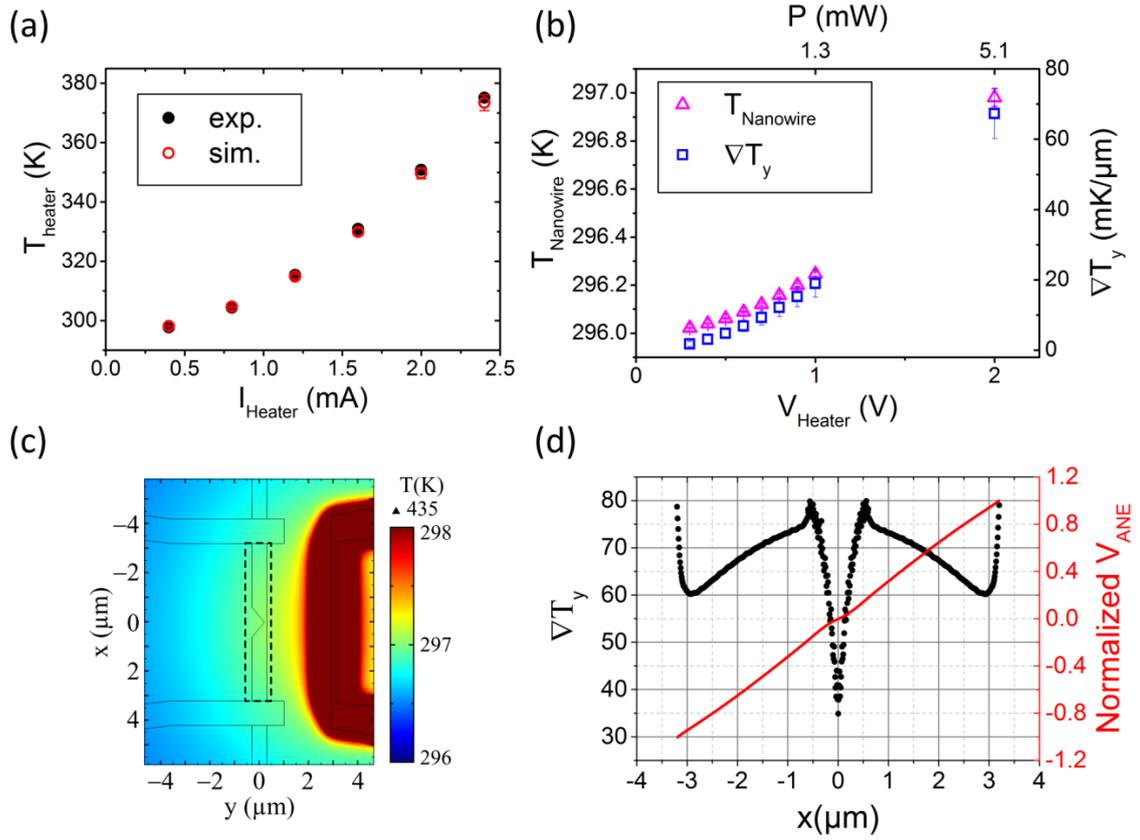

**Figure S4:** (a) Temperature of the heater from experiments (black dots) and simulations (red circles) as a function of heater current. (b) The dependence of temperature and average temperature gradient $\overline{\nabla T}_y$ of CoFeB nanowire on the heater voltage. (c) Contour map of the temperature distribution. (d) Local temperature gradient $\nabla T_y (x)$ (black dots) as a function of $x$ in the rectangular area indicated by dashed line in (c). Red line shows the sensitivity function of the normalized $V_{ANE}$ as a function of domain wall position

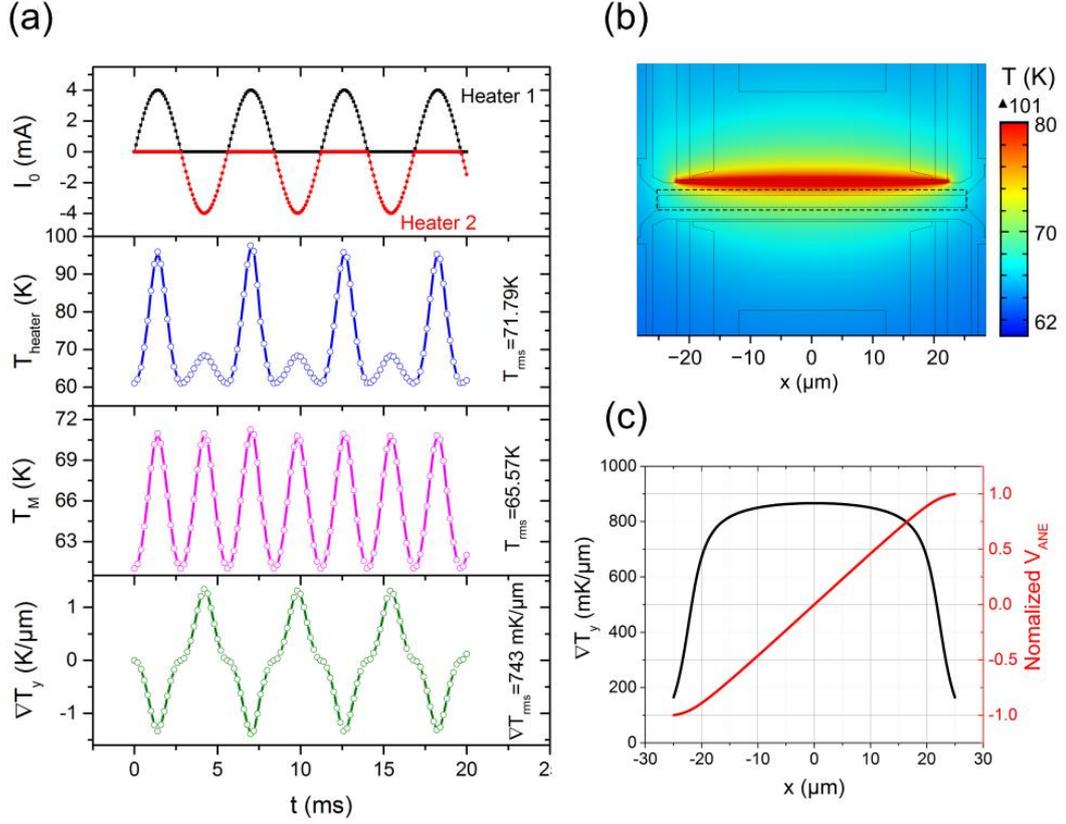

**Figure S5:** COMSOL simulation results of (GaMn)(AsP) nanowire. (a) Upmost panel shows the applied heater current in two heaters. Temperatures of one heater and nanowire varied with time are indicated by blue and magenta lines, respectively. The simulated temperature gradient across the wire (green line) with same frequency of the heater is indicated by the green line. (b) Contour map of the temperature distribution at 4.2 ms. (c) Local temperature gradient $\nabla T_y (x)$ (black line) as a function of $x$ in the rectangular area indicated by dashed line in (b). Red line shows the sensitivity function of the normalized $V_{ANE}$ as a function of domain wall position. In the range of -20 μm < $x_{DW}$ < 20 μm the sensitivity shows a dominantly linear behavior.

| Material | $c_p$ J/(kg K) | $\rho$ kg/m$^3$ | $\kappa$ W/(m·K) | $\rho_e$ S/ m |
|---|---|---|---|---|
| Silicon | [34]$C_{p\_Si}(T)$ | 2329 | [35]$\kappa_{\_Si}(T)$ | |
| Silica | [36]$C_{p\_SiO2}(T)$ | 2203 | [37]$\kappa_{\_SiO2}(T)$ | |
| Pt | [38]$C_{p\_Pt}(T)$ | 21450 | 71.6 | $\rho_e$ Pt(T) |
| Ta_Sample | 140 | 16400 | 57.5 | |
| [39]CoFeB | 440 | 8200 | 87 | |
| GaAs | [40]$C_{p\_GaAs}(T)$ | 5316 | [41]$\kappa_{\_GaAs}(T)$ | |
| (GaMn)(AsP) | [40]$C_{p\_GaAs}(T)$ | 5316 | [41]$\kappa_{\_GaAsP}(T)$ | |

**Table S1:** Parameters of the materials used in COMSOL simulation. $\rho_e$ was taken from the experimental data. Other data from literature as indicated in the references.

## I) ANE voltage calculation

In the following we derive the dependence of the ANE signal on the DW position for dominantly transverse in-plane gradients. Let us assume that the temperature gradient points in the y-direction as shown in Fig. S1. For simplicity we assume a linear temperature drop transversal to the nanowire, i.e., a constant local temperature gradient in y-direction $\nabla T_y(x)$ [see Fig. S1]. However we allow a variation of $\nabla T_y(x)$ along the x-direction as e.g. shown in Fig. S4(d). The electric field due to ANE at each position $x$ of the nanowire can be written as following:

$$\vec{E}_{ANE} = N_{ANE} \mu_0 \vec{M} \times \nabla T$$

where $\vec{M}$ and $\vec{\nabla T}$ are local magnetization and temperature gradient, respectively. The three components of the electric field are:

$$\begin{cases} E^x_{ANE} = N_{ANE}(\mu_0 M_y \nabla T_z - \mu_0 M_z \nabla T_y) \\ E^y_{ANE} = N_{ANE}(\mu_0 M_z \nabla T_x - \mu_0 M_x \nabla T_z) \\ E^z_{ANE} = N_{ANE}(\mu_0 M_x \nabla T_y - \mu_0 M_y \nabla T_x) \end{cases}$$

For the CoFeB nanowire with out-of-plane anisotropy we consider that the magnetization points only to the $z$ direction, i.e. $M_z = \pm M_S$ and $M_x = M_y = 0$, where $M_S$ is the saturation magnetization of the sample. For a small wire length element $dx$ as indicated in Fig. S1, we calculate the ANE voltage element as following:

$$dV_{ANE}(x) = -E^x_{ANE}(x)dx = N_{ANE} \mu_0 M_z(x) \nabla T_y(x) dx \quad (1)$$

With a single domain wall in the nanowire, as shown in Fig. S1, $M_z(x)$ equals $M_S$ on the interval $[-l/2, x_{DW}]$ and $-M_S$ on the interval $[x_{DW}, l/2]$. The total voltage between two contacts when the domain wall locates at $x = x_{DW}$ is:

$$V_{ANE}(x_{DW}) = -\int_{-\frac{l}{2}}^{\frac{l}{2}} dV_{ANE}(x) = -N_{ANE} \mu_0 M_S \left( \int_{-\frac{l}{2}}^{x_{DW}} \nabla T_y(x) dx - \int_{x_{DW}}^{\frac{l}{2}} \nabla T_y(x) dx \right) \quad (2)$$

If $\nabla T_y$ is constant between two contacts, the dependence of detected $V_{ANE}$ on the domain wall position can be written as:

$$V_{ANE}(x_{DW}) = -2N_{ANE} \mu_0 M_S \nabla T_y x_{DW} \quad (3)$$

We take the value at $x_{DW} = l/2$ as the maximum,

$$V_{ANE}^{max} = V_{ANE}\left(\frac{l}{2}\right) = -N_{ANE}\mu_0 M_S \int_{-l/2}^{l/2} \nabla T_y(x) dx = -N_{ANE}\mu_0 M_S l \overline{\nabla T}_y \qquad (4)$$

From Eq. (4) we can estimate the ANE coefficient $N_{ANE}$ based on the simulated average temperature gradient $\overline{\nabla T}_y$ derived from finite element modelling.

The detected $V_{ANE}$ is $x_{DW}$-dependent. Normalizing the $V_{ANE}(x_{DW})$, yields the sensitivity function:

$$\frac{V_{ANE}(x_{DW})}{V_{ANE}^{max}} = \frac{\int_{-\frac{l}{2}}^{x_{DW}} \nabla T_y dx - \int_{x_{DW}}^{\frac{l}{2}} \nabla T_y dx}{l \overline{\nabla T}_y} \qquad (5)$$

From Eq. (5) one can calculate the normalized ANE voltage by using a spatially varying $\nabla T_y(x)$. It allows to derive the DW position as function of the ANE voltage for a broad variety of heater and wire geometries. Figure S4(d) shows such derived normalized ANE signal as function of DW position along the CoFeB wires used in the experiments. Figure S5(c) shows the according signal for the (GaMn)(AsP) device.

**II)   ANE measurements on (GaMn)(AsP)**

Figure S2(a) illustrates the ANE measurement of DW propagation in the 25 nm thick (GaMn)(AsP) wire (yellow) of 50 μm length and 2 μm width. The temperature dependent resistance of the magnetic wire (Fig. S2(b)) allows to identify the time averaged magnetic wire temperature $T_M$. We adjust $T_M = 65.5$ K for all applied heater currents ranging from 0 up to 7 mA by regulating the cooling power of the cold-finger cryostat. The Curie temperature of our magnetic bar of Tc = 115 K is obtained by identifying the cusp[42] in dR/dT (Fig. S2(b)).

The 200 nm wide upper and lower Pt heater lines (red/blue) are fabricated by e-beam lithography and a lift-off procedure from a 50 nm thick Pt layer grown by e-beam evaporation. The heater current is generated by rectifying an alternating current in such a way that the positive (negative) half-wave is flowing along the upper (lower) heater generating a time-varying temperature gradient at the central bar which oscillates at the ac-current frequency of 123 Hz (Fig. S2(c)). The time-averaged heater temperature is deduced from measuring $V_{H1/H2}$ and comparing $V_{H1/H2}/I_0$ with the temperature dependency of the Pt heater line resistance.

The ANE signal is measured with a SR560 preamplifier followed by a SR830 lock-in amplifier. We have compared ANE signals measured at frequencies ranging from 18 Hz to 234 Hz without identifying a significant frequency dependency of the ANE signal. At higher frequencies, the signal amplitude becomes smaller. We conclude that the time-dependency of the thermal gradient at the bar follows the squared heater current at 123 Hz, with an additional sign change after each half period. This allows us to detect the ANE signal at the first harmonic of our heater current reference signal.

Figure S2(d) shows the quadratic dependence of the change of ANE signal as a function of the heater current amplitude $I_0$. The inset shows the ANE signal measured for heater currents of $I_0 = 1, 2, .. 7$ mA during a complete magnetisation reversal loop. The error bars in Fig. S2(d) are derived from the standard deviation of the ANE signal measured during the magnetisation reversal loops.

We attribute the higher coercive field at larger heater current (see inset to Fig. S2(d)) to DW nucleation in the wider contact area outside the magnetic wire. Note, that larger cooling is required to keep the magnetic wire temperature equal for all heater currents. Therefore, the contact temperature is lower at higher heater currents. Considering the thermal activation of the nucleation process explains the observed temperature dependence of the coercive field.

To complement the data on spin torque induced DW propagation in (GaMn)(AsP) in Fig. 3(c) of the main manuscript, Fig. S3 shows the backward propagation of the DW by current pulses of inverted polarity. Here the DW was first nucleated by a nucleation pulse and then propagated to the positive x-direction (right hand side of the wire) by a series of current pulses. Then the pulse polarity was inverted and the backward propagation was detected by ANE measurements.

### III) COMSOL simulations and thermal calibrations

The temperature gradients across the CoFeB and (GaMn)(AsP) nanowires were estimated from COMSOL simulations by using a Joule heating package. Geometries of the modeled samples were taken from the experimental design. Lateral simulated size is 1mm×1mm. Bottom temperature of the substrate (Si or GaAs) was set to the ambient temperature. In the simulations we took temperature dependent parameters such as thermal conductivity $\kappa$ and heat capacity $c_p$ from references [1-8], as listed in the Table S1. Note that these values are for bulk materials apart from the electrical resistivity of the Pt heater $\rho_e$ which comes from experimental data. When the materials are in form of thin film, the thermal conductivity might be lower due to additional interfaces. Also impurities from sample fabrication could lower the thermal conductivity. Our simulations thus provide low boundaries of temperature gradients.

To validate the simulation parameters calibration samples were characterized and simulated. The heater lines were heated by an electrical current $I_{\text{Heater}}$. The applied power $P_{\text{heat}}$ was derived from the measured heater line resistance. The temperature $T_{\text{heater}}$ of the heater line during heating was derived from the change of resistance $R_{\text{Pt}}(T_{\text{heater}})$ where $R_{\text{Pt}}(T_{\text{heater}})$ was calibrated using a variable temperature probe station. Fig S4(a) shows the calculated and simulated temperature increase of the heater line as function of the applied heater current for a typical calibration sample. The black dots show the experimental data from the temperature calibration. The simulation results (red circles) are in good agreement with the experimental ones demonstrating the general validity of the simulation parameters.

Based on these simulation parameters, Fig. S4(b) shows the simulated temperature increase and the average temperature gradient $\overline{\nabla T}_y$ of the CoFeB nanowire (i.e. $\nabla T_y(x)$ averaged over the wire length) as function of the heater voltage. For a voltage of 2V (or power of 5.1 mW) applied to the heater line we obtain a $\overline{\nabla T}_y$ of 67 ±7 mK/µm. The uncertainty is derived from the standard deviation of the difference of $\nabla T_y(x)$ and $\overline{\nabla T}_y$. Figure S4(c) shows a typical temperature distribution on top of the sample. For the area indicated by the dashed rectangular we plot the temperature gradient $\nabla T_y(x)$ as a function of $x$ as indicated by black dots in Fig. S4(d). Significant variations of $\nabla T_y(x)$ over the wire length are found near the notch ($x = 0$) and near the electrical contacts ($x = \pm\ 3.2$ µm). Near the notch the heat flux is locally redirected into the wider part of the wire ($y$-direction) leading to a drop of $\nabla T_y(x)$. In contrast near the contacts an increase of $\nabla T_y(x)$ is found. The superimposed variation (lower gradient near the contacts, higher gradient in the center) is due to the limited length of the heater line compared to the wire segment.

Using the local $\nabla T_y(x)$ and Eq. (2) we calculate a normalized ANE signal between two contacts as a function of the domain wall position. The resulting sensitivity function is indicated by the red line in Fig. S4(d). Note that despite the local variations of $\nabla T_y(x)$ the sensitivity function can still be reasonably approximated by a straight line i.e. by the approximated sensitivity of Eq. (2) of the main paper.

In the simulation for (GaMn)(AsP) we use a time-dependent model to simulate the effect of the double heater scheme as described in II. Figure S5(a) shows from top to bottom: the heater currents through the two heaters (red, black), the temperature $T_{\text{heater}}$ of heater 1, the temperature of the magnetic wire $T_M$, and the average temperature gradient $\nabla T_y$ across the nanowire as a function of time.

The wire temperature $T_M$ oscillates with the double frequency $2f$ of the heater frequency $f$. The gradient shows a dominant oscillation at $f$. The temperature of heater 1 (blue) shows a strong temperature increase during self-heating and a weaker temperature increase while heater 2 is heated by a current. For a heater current $I_{heater}$ = 4mA, heater and nanowire temperatures match well with the experimental data. For these parameters the simulations yield a temperature gradient of $\overline{\nabla T}_y$ = 743 ± 187 mK /µm averaged over the whole wire length.

**References:**

34 A.S. Okhotin, A.S. Pushkarskii, V.V. Gorbacher, Thermophysical properties of semiconductors, Moscow, "Atom" Publ. House, 1972 (Russian).
35 C.J. Glassbrenner, G A. Slack, *Phys. Rev.* **134,** A1058 (1964).
36 R.P. Sosman, The properties of silica, Reinhold, Newyork, 1927.
37 Y.S. Touloukian, R.W. Powell, C.Y. Ho, P.G. Klemens, Thermophysical properties of matter, Vol. 2, IFI/Plenum, New York, p.193 (1970).
38 G.T. Furukawa, M.L. Reilly, J.S. Gallagher, J. Phys. Chem. Ref. Data, **3**, 163(1974)
39 M. Walter, J. Walowski, V. Zbarsky, M. Münzenberg, M. Schäfers, D. Ebke, G. Reiss, A. Thomas, P. Peretzki, M. Seibt, J.S. Moodera, M. Czerner, M. Bachmann, C. Heiliger, Nature Mater. **10**, 742 (2011).
40 R.O. Carlson, G.A. Slack, S.J. Silverman, J. Appl. Phys. **36**, 505 (1965).
41 J.S. Blakemore, J. Appl. Phys. **53**, R123 (1982).
42 V. Novak, et al. , Phys. Rev. Lett. **101**, 077201 (2008).